\documentclass[sigconf]{acmart}
\AtBeginDocument{%
  }

\setcopyright{none}
\copyrightyear{2018}
\acmYear{2018}
\acmDOI{XXXXXXX.XXXXXXX}
\acmConference[Conference acronym 'XX]{Make sure to enter the correct
  conference title from your rights confirmation email}{June 03--05,
  2018}{Woodstock, NY}
\acmISBN{978-1-4503-XXXX-X/2018/06}
\usepackage{algorithm}
\usepackage{algorithmic}
\usepackage{booktabs}
\usepackage{longtable}
\usepackage{bm}
\usepackage{xcolor}
\usepackage{multirow}
\usepackage{graphicx}
\usepackage{makecell}
\usepackage{subcaption}
\usepackage{lipsum}
\usepackage{enumitem}
\usepackage{amsmath}
\usepackage{amsfonts}
\usepackage{threeparttable}
\usepackage{graphicx}
\usepackage{subcaption} 

\usepackage{amssymb}

\begin{document}

\title{DTRec: Learning Dynamic Reasoning Trajectories for Sequential Recommendation}

\author{Yifan Shao}
\authornote{Both authors contributed equally to this research.}
\affiliation{%
  \institution{The Chinese University of Hong Kong}
  \city{Hong Kong}
  \country{China}
  }
\email{yifashao209@gmail.com}

\author{Peilin Zhou}
\authornotemark[1]

\affiliation{%
  \institution{Hong Kong University of Science and Technology (Guangzhou)}
  \city{Guangzhou}
  \country{China}
}
\email{zhoupalin@gmail.com}

\author{Shoujin Wang}
\affiliation{%
  \institution{University of Technology Sydney}
  \city{Sydney}
  \country{Australia}}
\email{shoujin.wang@uts.edu.au}

\author{Weizhi Zhang}
\affiliation{%
  \institution{University of Illinois Chicago}
  \city{Chicago}
  \state{Illinois}
  \country{USA}
}
\email{wzhan42@uic.edu}

\author{Xu Cai}
\affiliation{%
 \institution{Jilin University}
 \city{Changchun}
 \country{China}
 }
\email{caixu5522@mails.jlu.edu.cn}

\author{Sunghun Kim}
\affiliation{%
 \institution{Hong Kong University of Science and Technology}
 \city{Hong Kong}
 \country{China}
 }
\email{hunkim@ust.hk}

\renewcommand{\shortauthors}{Shao et al.}


\begin{abstract}
Inspired by advances in LLMs, reasoning-enhanced sequential recommendation performs multi-step deliberation before making final predictions, unlocking greater potential for capturing user preferences. However, current methods are constrained by static reasoning trajectories that are ill-suited for the diverse complexity of user behaviors. They suffer from two key limitations: (1) a static reasoning \textbf{direction}, which uses flat supervision signals misaligned with human-like hierarchical reasoning, and (2) a fixed reasoning \textbf{depth}, which inefficiently applies the same computational effort to all users, regardless of pattern complexity. These rigidity lead to suboptimal performance and significant computational waste. 

To overcome these challenges, we propose \textbf{DTRec}, a novel and effective framework that explores the \textbf{D}ynamic reasoning \textbf{T}rajectory for Sequential \textbf{R}\textbf{ec}ommendation along both direction and depth. To guide the \textbf{direction}, we develop Hierarchical Process Supervision (HPS), which provides coarse-to-fine supervisory signals to emulate the natural, progressive refinement of human cognitive processes. To optimize the \textbf{depth}, we introduce the Adaptive Reasoning Halting (ARH) mechanism that dynamically adjusts the number of reasoning steps by jointly monitoring three indicators. Extensive experiments on three real-world datasets demonstrate the superiority of our approach, achieving up to a 24.5\% performance improvement over strong baselines while simultaneously reducing computational cost by up to 41.6\%. 
\end{abstract}



\maketitle
\section{Introduction}
Sequential recommendation aims to predict the next item which a user will interact with based on their historical behavior sequence. In recent years, neural sequential recommenders such as SASRec~\cite{kang2018self} and BERT4Rec~\cite{sun2019bert4rec} have achieved remarkable progress by leveraging Transformer-based architecture to model sequential dependencies. However, these approaches lack a deliberate reasoning process, limiting the ability to understand the complex evolving nature of user preferences.

Motivated by chain-of-thought (CoT) prompting in large language models (LLMs), recent studies have explored reasoning-enhanced sequential recommendation~\cite{tang2025thinkrecommendunleashinglatent,liu2025lareslatentreasoningsequential} to perform multi-step reasoning before generating the final prediction. By introducing intermediate reasoning steps, these methods iteratively refine the understanding of user intent, thereby enhancing both the accuracy and interpretability of recommendations. Despite promising results, existing approaches still learn \textbf{\emph{static reasoning trajectories}}, which lack the adaptability to the diverse and complex patterns found in real-world user behavior\cite{cen2020controllable}. Specifically, this static nature appears in two aspects: (1) \textbf{static reasoning direction}: by supervising all intermediate steps with the final target item, they create a ``flat'' process supervision signal. This approach fundamentally misaligns with the hierarchical, coarse-to-fine nature of human cognition\cite{navon1977forest}, which progresses from broad overviews (\textit{e.g.}, product category) to specific details (\textit{e.g.}, item's attributes like brand or color). (2) \textbf{static reasoning depth}: the number of reasoning steps is fixed for all samples, regardless of their varying complexity (\textit{e.g.}, a simple, predictable pattern versus a sequence with abrupt shifts in interest). This leads to misallocation of computational resources, where simple cases are over-processed while complex ones are under-reasoned.

To address these limitations, we propose \textbf{DTRec}, a framework that explores the \textbf{D}ynamic reasoning \textbf{T}rajectory for Sequential \textbf{R}\textbf{ec}ommendation along two complementary dimensions: \textit{direction} and \textit{depth}. For \textit{dynamic reasoning direction}, we introduce \textbf{Hierarchical Process Supervision (HPS)}, which aligns the supervision signal with the abstraction level of each reasoning step. Specifically, HPS constructs multi-level semantic prototypes via K-means clustering over item embeddings to guide a coarse-to-fine reasoning process: early steps are supervised by coarse-grained prototypes (broad categories), while later steps are supervised by fine-grained ones (specific attributes), shaping more structured and informative reasoning trajectories. For \textit{dynamic reasoning depth}, we propose an \textbf{Adaptive Reasoning Halting (ARH)} mechanism that dynamically determines the number of reasoning steps. ARH jointly monitors prediction confidence, inter-step output consistency, and representation stability to decide when to terminate reasoning, enabling more efficient computation allocation. Our main contributions can be summarized as follows:
\begin{itemize}
[left=0pt]
\item We introduce DTRec, a dynamic reasoning framework that adaptively adjusts both reasoning direction and reasoning depth for sequential recommendation.

\item We propose Hierarchical Process Supervision to provide coarse-to-fine, step-aware supervision over the reasoning trajectory, and Adaptive Reasoning Halting to dynamically control reasoning depth, improving efficiency without compromising performance.

\item Extensive experiments on three real-world datasets demonstrate the effectiveness of our approach, achieving up to 24.5\% performance improvement over strong baselines while reducing computational cost by 30\%.
\end{itemize}

\section{Preliminary}
\subsection{Problem Definition}
For user $u$, we define their chronological interaction sequence as $\mathcal{S}^u=[i_1, i_2, \ldots, i_{n}]$. Instead of directly predicting the next item, reasoning-enhanced sequential recommendation introduces an intermediate reasoning sequence $R^u = \{r_1, r_2, \cdots, r_T\}$. The final recommendation is then generated conditioned on both the original user behavior and this reasoning sequence:
\begin{equation}
    P(\hat{i}_{n+1}\mid S^u) =  P(\hat{i}_{n+1} \mid R^u, S^u)\cdot P(R^u \mid S^u) .
\end{equation}
\subsection{Process supervision}
To provide process-level supervision, ReaRec~\cite{tang2025thinkrecommendunleashinglatent} requires each reasoning state $r_t$ to directly predict the ground-truth target item $v_\star$. The process loss $\mathcal{L}_{\text{0}}$ is the sum of the cross-entropy losses over all steps:
\begin{equation}
\mathcal{L}_{\text{0}} = -\sum_{t=0}^{T} \log \hat{y}^{(t)}_{v_\star},
\end{equation}
where the prediction probability $\hat{y}^{(t)}$ is calculated as $\text{softmax}(r_t \cdot \mathbf{E}^\top)$ using the item embedding table $\mathbf{E}$.

\begin{figure}[h]
    \centering
    \includegraphics[width=0.5\textwidth]{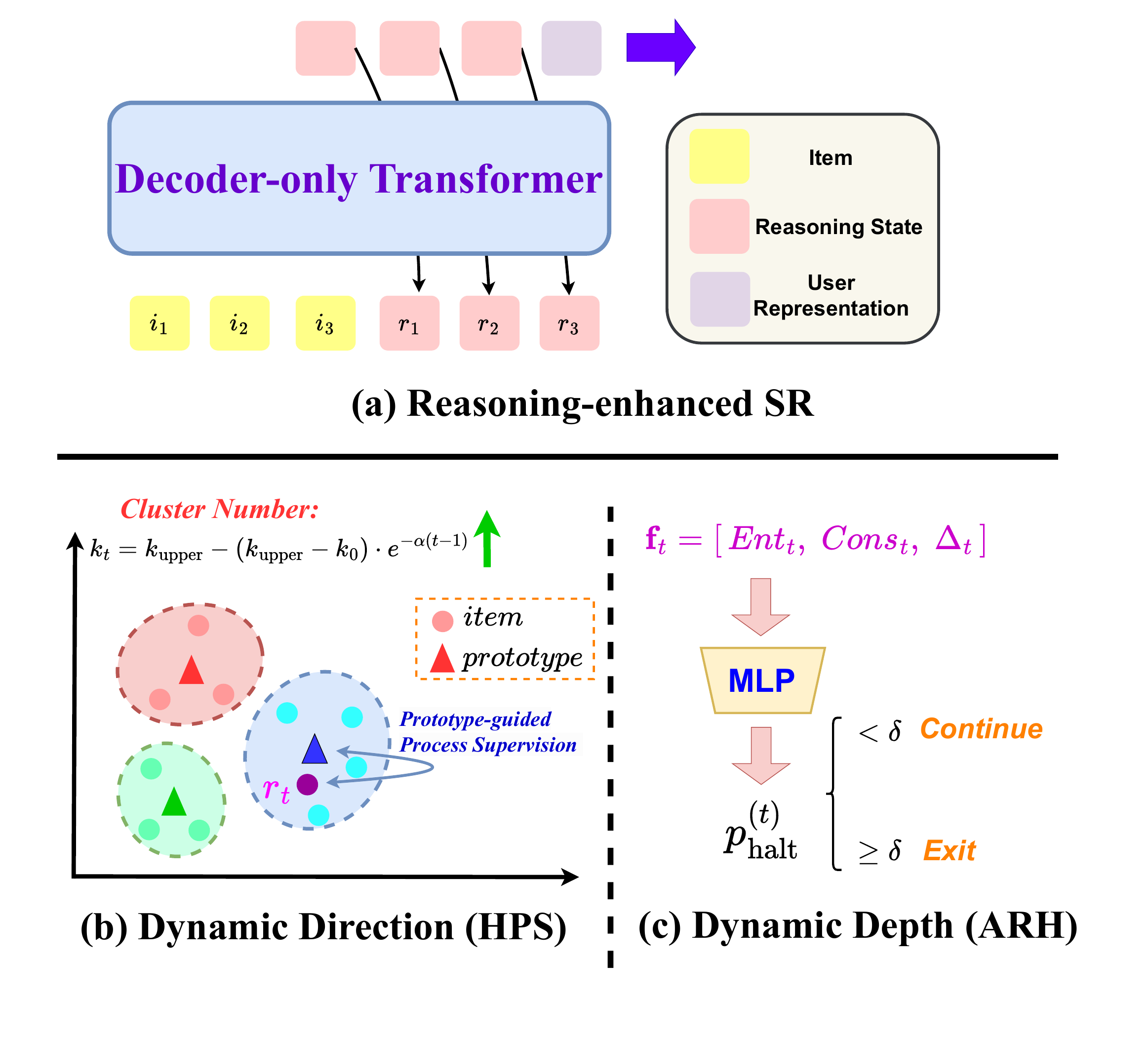}
    \vspace{-0.5in}
    \caption{The overall architecture of our DTRec framework.}
    \label{figure:overview}
\end{figure}

\section{Methodology}
In this section, we introduce the proposed DTRec, which consists of two key components: Hierarchical Process Supervision (HPS) and Adaptive Reasoning Halting (ARH). The overall architecture DTRec is illustrated in Figure~\ref{figure:overview}.

\subsection{Hierarchical Process Supervision (HPS)}\label{section-Process_Supervision}
Human-style reasoning naturally follows a \textbf{\textit{coarse-to-fine}} pattern. For instance, while the ultimate prediction target is a specific item (\emph{e.g.}, ``iPhone 15 Pro"), human-like reasoning may first identify the broad category (``electronics"), then narrow to the subcategory (``smartphones"), and finally pinpoint specific attributes (``Apple", ``256GB", etc.)
However, existing process supervision methods~\cite{tang2025thinkrecommendunleashinglatent} directly supervise each intermediate reasoning state with the target item. This overly strong signal does not align with the progressive nature of human reasoning and may lead to incorrect reasoning trajectory directions.  
To address this limitation, we propose \textbf{Hierarchical Process Supervision (HPS)} that dynamically adjusts the granularity of process signals across reasoning steps.

\subsubsection{Semantic Prototype Extraction} 

We apply K-means clustering to the item embedding table. Since items 
with similar features cluster together in the embedding space, the 
resulting cluster centers naturally serve as semantic prototypes that 
represent shared abstract attributes such as categories, brands, and styles.

\begin{equation}
\{c_i^{(t)}\}_{i=1}^{k_t} = \text{K-means}(\mathbf{E}, k_t),
\end{equation}
where $k_t$ is the number of clusters at reasoning step $t$. Critically, instead of using a fixed $k$, we progressively increase the granularity of prototype to enable coarse-to-fine reasoning:
\begin{equation}
k_t = k_{\text{upper}} - (k_{\text{upper}} - k_0) \cdot e^{-\alpha(t-1)},
\end{equation}
where $k_0$ is the initial cluster number, $\alpha > 0$ controls the expansion rate, and $k_{\text{upper}}$ prevents over-fragmentation.

\subsubsection{Prototype-guided Process Supervision}
At each reasoning step $t$, we retrieve the corresponding semantic prototype $p_t$ for reasoning state $r_t$ by finding the nearest cluster center:
\begin{equation}
p_t = \mathop{\arg\min}_{c_i^{(t)}} \|r_t - c_i^{(t)}\|_2, \quad i \in \{1,2,\dots,k_t\}.
\end{equation}
The prototype $p_t$ represents the abstract semantic attributes reasoned at this step, serving as a \textbf{\textit{soft target}} to guide the learning of $r_t$. We first compute recommendation probabilities $\hat{y}^{(t)}$ and $\hat{y}_p^{(t)}$ for $r_t$ and $p_t$ respectively, then apply cross-entropy loss to align their representations:
\begin{equation}
\hat{y}^{(t)} = \text{softmax}(r_t \cdot \mathbf{E}^\top), \quad 
\hat{y}_p^{(t)} = \text{softmax}(p_t \cdot \mathbf{E}^\top),
\end{equation}
\begin{equation}
\mathcal{L}_{p} = - \sum_{t=0}^{T}  \hat{y}_p^{(t)} \cdot \log \hat{y}^{(t)}.
\end{equation}
The final training objective is a weighted sum of the standard process loss $\mathcal{L}_{\text{0}}$ and our prototype-guided process loss $\mathcal{L}_{p}$. Note that we adopt a warm-up strategy, increasing the weight of $\mathcal{L}_{p}$ from 0 to its full value in the first 10 epochs to avoid unstable supervision caused by undertrained item embeddings in the early stage.

\subsection{Adaptive Reasoning Halting (ARH)}\label{section-Adaptive_Halting}

In real-world scenarios, user behavior patterns vary widely in complexity. Applying a fixed reasoning depth to all sequences can lead to unnecessary computation on simple patterns and insufficient reasoning on complex ones. To overcome this shortcoming, we propose \textbf{Adaptive Reasoning Halting (ARH)}, which dynamically adjusts the reasoning depth based on current reasoning states. At reasoning step $t$, ARH extracts three complementary indicators that together capture the convergence status of the reasoning process:

\begin{itemize}[leftmargin=*, topsep=2pt, itemsep=2pt]
    \item \textbf{Prediction Entropy ($Ent_t$):} Calculated as $- \sum_{i} \hat{y}^{(t)}_i \log \hat{y}^{(t)}_i$, qu-\\antifying the uncertainty of the current prediction.
    
    \item \textbf{Inter-step Consistency ($Cons_t$):} Defined by $D_{KL}(\hat{y}^{(t-1)} \Vert \hat{y}^{(t)})$, measuring the consistency of the model's output. 
    
    \item \textbf{Representation Variation ($\Delta_t$)} Given by $ \| r_t - r_{t-1} \|_2$, tracking the internal stability of the hidden reasoning states.
\end{itemize}

To fuse these diverse signals into a single criterion, we first concatenate these indicators into a comprehensive feature vector $\mathbf{f}_t$. This vector is then processed by a lightweight Multi-Layer Perceptron (MLP) to compute the final halting probability $p_{\mathrm{halt}}^{(t)}$:
\begin{equation}
\mathbf{f}_t = [\,Ent_t,\; Cons_t,\; \Delta_t\,],
\end{equation}
\begin{equation}
p_{\mathrm{halt}}^{(t)} = \mathrm{MLP}(\mathbf{f}_t),
\end{equation}

\subsubsection{Training} 
To avoid the non-differentiability of discrete halting decisions, we adopt a \textit{soft halting} training scheme. The final prediction $\hat{y}$ is computed as a weighted sum of step-wise predictions:
\begin{equation}
\hat{y} = \sum_{t=1}^{T} w_t \hat{y}^{(t)}, \quad w_t = p_{\mathrm{halt}}^{(t)} \prod_{j<t} (1-p_{\mathrm{halt}}^{(j)}),
\end{equation}
where $w_t$ denotes the contribution of step $t$ to the final prediction.

\subsubsection{Inference} 
During inference, we perform \textit{discrete early exit}: reasoning halts at the step $t$ when $p_{\mathrm{halt}}^{(t)}$ exceeds a predefined threshold $\delta$.

\section{Experiments}
\subsection{Experimental Setup}
\subsubsection{Datasets and Evaluation Metrics}
We conduct extensive experiments on three real-world recommendation datasets: Sports, Beauty and Yelp. The detailed statistics of the datasets are summarized in Table~\ref{table:dataset}. For a fair comparison, we follow the same data preprocessing and evaluation metrics as in previous work~\cite{zhou2024equivariantcontrastivelearningsequential}.

\subsubsection{Compared Models.}
Considering the proposed DTRec is model-agnostic and plug-and-play, we selected several representative base models to compare their performance with and without our method, including SASRec~\cite{kang2018self}, GRU4Rec~\cite{hidasi2015session} and BERT4Rec~\cite{sun2019bert4rec}. We also conducted comparisons with the existing reasoning-enhanced model ReaRec~\cite{tang2025thinkrecommendunleashinglatent} and LARES~\cite{liu2025lareslatentreasoningsequential}.

\subsubsection{Implementation Details}
For a fair comparison, the hyperparameter settings for all baselines are adopted from their original papers. For the proposed DTRec, the maximum cluster number $k_{upper}$ is set as 3000. We carefully tune the minimum cluster number $k_{0}$ in $[10,100]$, the granularity expansion rate $\alpha$ in $(0,1)$, the weight of $\mathcal{L}_{p}$ in $\{10^{-2}, 10^{-1}, 1\}$, and the halting threshold $\delta$ in $[0.3, 0.8]$.

\begin{table}[!t]
    \caption{Statistics of the datasets}
    \vspace{-0.15in}
	\begin{tabular}{c *{4}{r}}
		\toprule
		\textbf{Datasets} & \textbf{\#Users} & \textbf{\#Items} & \textbf{\#Interactions} & \textbf{Sparsity}\\
		\midrule
		Sports 	& 355,98  & 18,357  & 296,337   & 99.95\%  \\
		Beauty 	& 22,363 & 12,101 & 198,502 & 99.93\% \\
		Yelp 	& 30,499 & 20,068 & 317,182 & 99.95\% \\
		\bottomrule
	\end{tabular}
    \label{table:dataset}
\end{table}

\begin{table*}[t]
\caption{Performance comparison of different models on three datasets. The best and second-best results are indicated in bold and underlined font. $^*$ indicates the statistical significance for $p$ \textless 0.01 compared to the best baseline.}
\vspace{-0.1in}
\centering
\tabcolsep=0.07cm
\renewcommand{\arraystretch}{1.2}
\resizebox{\textwidth}{!}{
\begin{tabular}{ll|cccc|cccc|cccc}
\toprule[1pt]
\multirow{2}{*}{\textbf{Backbone}} & \multirow{2}{*}{\textbf{Model}} & \multicolumn{4}{c|}{\textbf{Sports}} & \multicolumn{4}{c|}{\textbf{Beauty}} & \multicolumn{4}{c}{\textbf{Yelp}} \\ 
\cmidrule{3-14} 
&  & Recall@10 & NDCG@10 & Recall@20 & NDCG@20 & Recall@10 & NDCG@10 & Recall@20 & NDCG@20 & Recall@10 & NDCG@10 & Recall@20 & NDCG@20 \\ 
\midrule
\multirow{4}{*}{SASRec} 
& - Base & 0.0445 & 0.0212 & 0.0692 & 0.0274 & 0.0664 & 0.0316 & 0.1044 & 0.0411 & 0.0605 & 0.0381 & 0.0868 & 0.0448 \\
& - ReaRec & 0.0475 & 0.0224 & \underline{0.0742} & \underline{0.0293} & 0.0686 & 0.0328 & 0.1071 & 0.0416 & 0.0621 & 0.0385 & 0.0895 & 0.0455 \\
& - LARES & \underline{0.0481} & \underline{0.0230} & 0.0728 & 0.0292 & \underline{0.0707} & \underline{0.0332} & \underline{0.1100} & \underline{0.0432} & \underline{0.0627} & \underline{0.0390} & \underline{0.0903} & \underline{0.0459} \\
& - DTRec & \textbf{0.0553}* & \textbf{0.0277}* & \textbf{0.0855}* & \textbf{0.0354}* & \textbf{0.0828}* & \textbf{0.0409}* & \textbf{0.1294}* & \textbf{0.0527}* & \textbf{0.0682}* & \textbf{0.0407}* & \textbf{0.0992}* & \textbf{0.0476}* \\ 
\midrule
\multirow{4}{*}{GRU4Rec} 
& - Base & 0.0342 & 0.0175 & 0.0549 & 0.0227 & 0.0569 & 0.0281 & 0.0871 & 0.0357 & 0.0397 & 0.0200 & 0.0651 & 0.0264 \\
& - ReaRec & \underline{0.0365} & \underline{0.0181} & \underline{0.0573} & \underline{0.0237} & 0.0582 & 0.0286 & 0.0901 & 0.0366 & \underline{0.0410} & 0.0211 & 0.0669 & 0.0273 \\
& - LARES & 0.0351 & 0.0180 & 0.0569 & 0.0231 & \underline{0.0605} & \underline{0.0308} & \underline{0.0965} & \underline{0.0399} & 0.0404 & \underline{0.0212} & \underline{0.0675} & \underline{0.0280} \\
& - DTRec & \textbf{0.0402}* & \textbf{0.0208}* & \textbf{0.0642}* & \textbf{0.0265}* & \textbf{0.0677}* & \textbf{0.0338}* & \textbf{0.1037}* & \textbf{0.0432}* & \textbf{0.0451}* & \textbf{0.0232}* & \textbf{0.0733}* & \textbf{0.0296}* \\ 
\midrule
\multirow{4}{*}{BERT4Rec} 
& - Base & 0.0244 & 0.0124 & 0.0394 & 0.0162 & 0.0394 & 0.0188 & 0.0651 & 0.0253 & 0.0274 & 0.0136 & 0.0470 & 0.0185 \\
& - ReaRec & \underline{0.0301} & \underline{0.0151} & \underline{0.0459} & \underline{0.0175} & 0.0427 & 0.0194 & 0.0678 & 0.0271 & \underline{0.0306} & \underline{0.0151} & 0.0509 & 0.0197 \\
& - LARES & 0.0274 & 0.0131 & 0.0432 & 0.0172 & \underline{0.0481} & \underline{0.0213} & \underline{0.0772} & \underline{0.0277} & 0.0294 & 0.0144 & \underline{0.0528} & \underline{0.0201} \\
& - DTRec & \textbf{0.0357}* & \textbf{0.0173}* & \textbf{0.0552}* & \textbf{0.0232}* & \textbf{0.0554}* & \textbf{0.0272}* & \textbf{0.0874}* & \textbf{0.0352}* & \textbf{0.0406}* & \textbf{0.0207}* & \textbf{0.0665}* & \textbf{0.0226}* \\ 
\bottomrule[1pt]
\end{tabular}
}
\label{table:overall}
\end{table*}

\subsection{Overall Performance}
As shown in Table~\ref{table:overall}, SASRec is the top-performing baseline. All reasoning-enhanced frameworks deliver substantial improvements upon it, confirming the value of the think-before-action paradigm. Among them, our proposed DTRec particularly excels, comprehensively outperforming advanced models like ReaRec and LARES. This superiority can be attributed to the coarse-to-fine reasoning direction guided by HPS and the adaptive reasoning depth enabled by ARH.

\begin{table}
\caption{
Ablation study of DTRec. “N@10” denotes NDCG@10, and “Cons.” indicates the relative computational cost, which is calculated as the quotient of average reasoning steps with and without early exit.
}
\vspace{-0.15in}
\centering
\resizebox{0.45\textwidth}{!}{ 
\setlength{\tabcolsep}{0.5em} 
\renewcommand\arraystretch{1} 
\begin{tabular}{lcccc}
\toprule
\multirow{2}{*}{\textbf{Model}} & \multicolumn{2}{c}{\textbf{Sports}} & \multicolumn{2}{c}{\textbf{Beauty}} \\
\cmidrule(lr){2-3} \cmidrule(lr){4-5}
& \textbf{N@10} & \textbf{Cons.} & \textbf{N@10} & \textbf{Cons.} \\ 
\midrule 
  \enspace Base     & 0.0212          & 100\%   & 0.0316          & 100\%   \\
 +HPS (k=const)     & 0.0248          & 100\%   & 0.0362          & 100\%   \\
 +HPS (w/o warmup)  & 0.0258          & 100\%   & 0.0392          & 100\%   \\
 +HPS               & 0.0265          & 100\%   & 0.0398          & 100\%   \\
 +HPS+REE           & 0.0269          & 85.6\%   & 0.0407         & 67.1\%   \\
 +HPS+ARH (\textbf{ours})    & \textbf{0.0277}          & \textbf{69.2}\%   & \textbf{0.0409}         & \textbf{58.4\%}   \\
\bottomrule
\end{tabular}}

\label{table:ablation}
\end{table}

\subsection{Ablation Study}
The ablation study in Table~\ref{table:ablation} validates the effectiveness of our key components. First, the results highlight the effectiveness of HPS's coarse-to-fine supervision. A variant with a fixed cluster number (k=const) yields only limited gains, while removing the warm-up phase harms training stability by introducing noise from undertrained embeddings. Second, our ARH mechanism proves superior to a simpler Representation-based Early Exit (REE), which directly trains a binary halting head on the reasoning state $r_t$. We believe this more reliable halting decision stems from the fusion of multiple indicators.

\begin{figure}[t]
    \centering 
    \begin{subfigure}[t]{0.47\linewidth}
        \centering
        \includegraphics[width=1\textwidth]{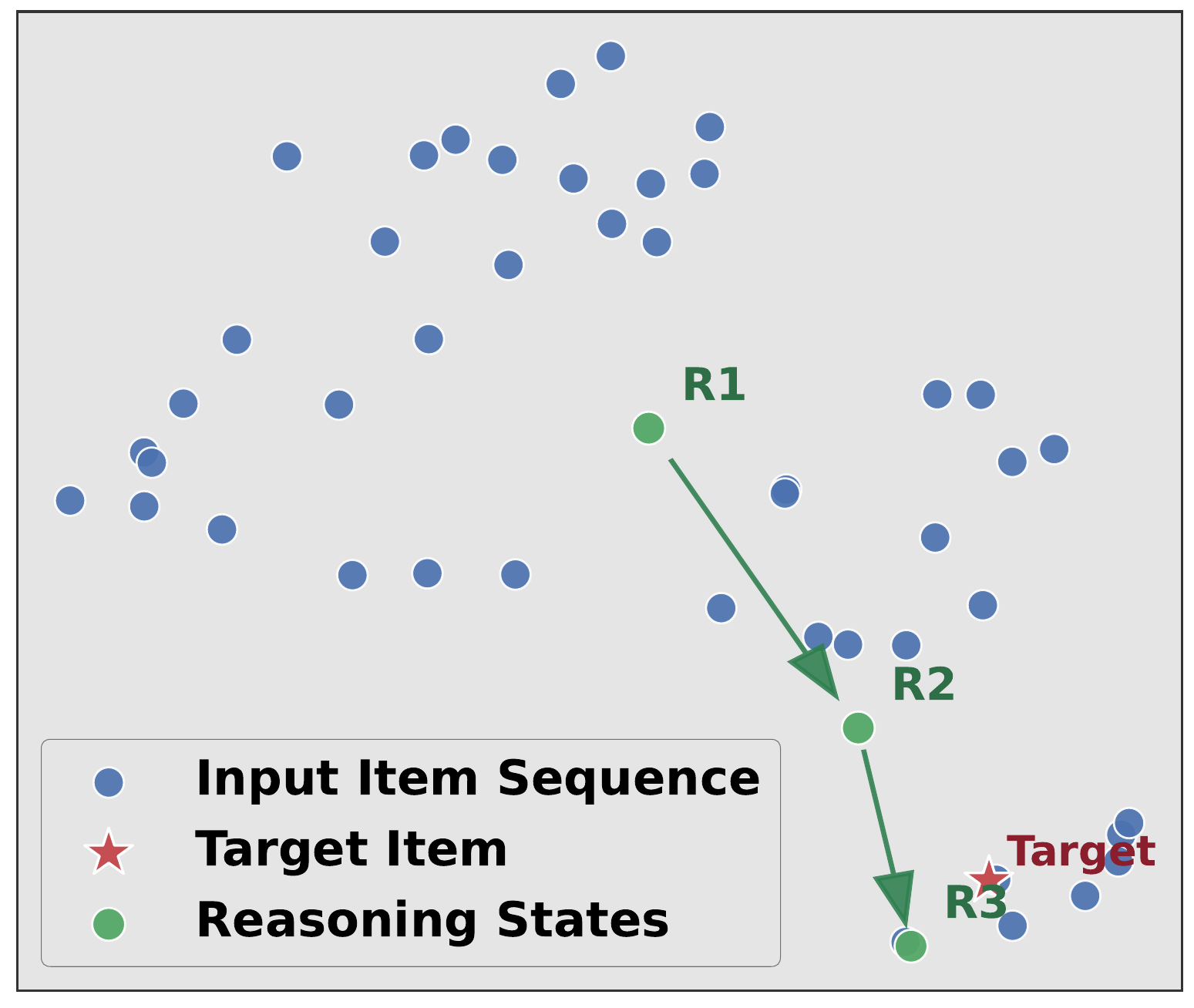}
        \caption{DTRec} 
        \label{fig:dtrec} 
    \end{subfigure}
    \hfill 
    \begin{subfigure}[t]{0.47\linewidth}
        \centering
        \includegraphics[width=1\textwidth]{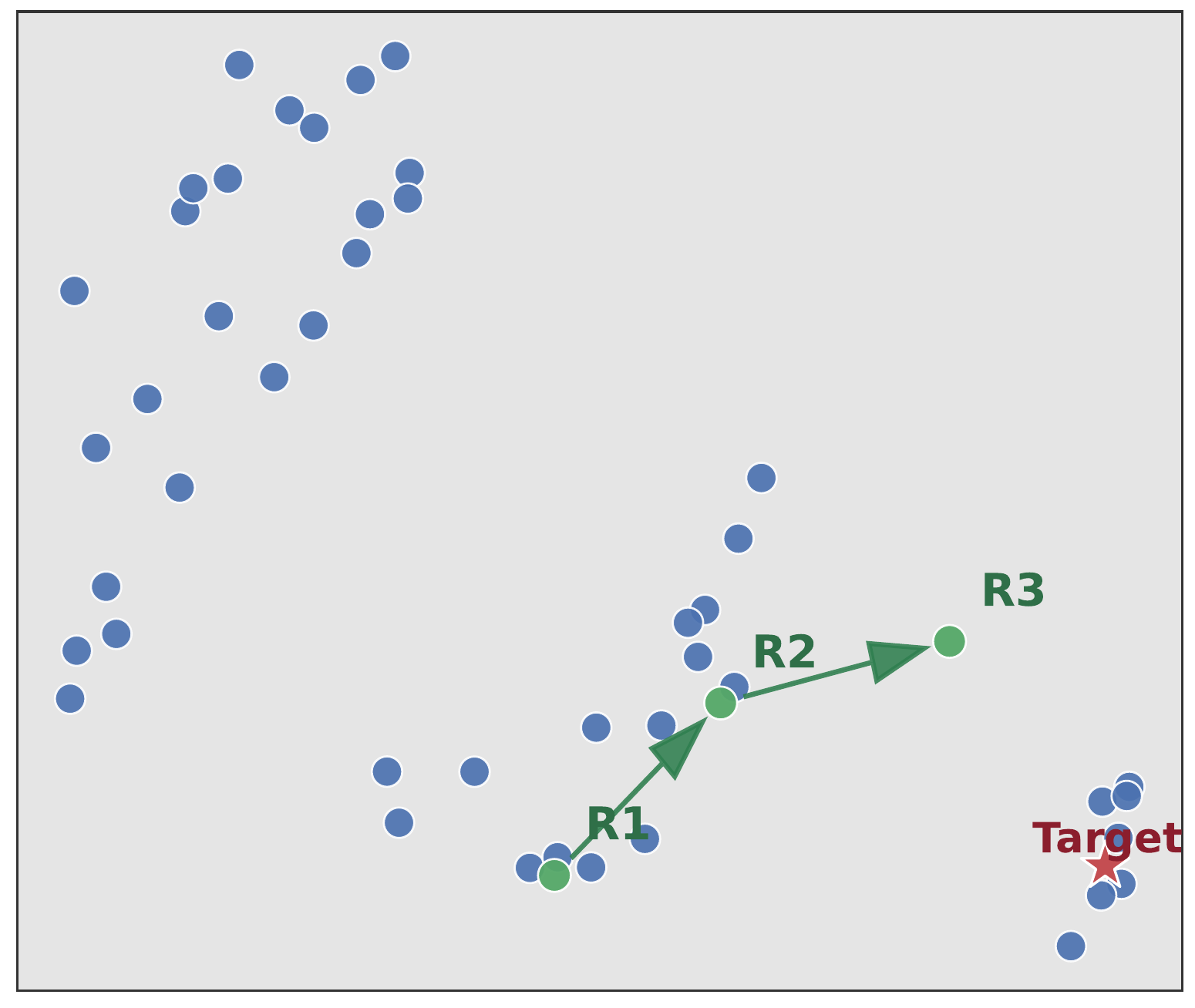}
        \caption{ReaRec} 
        \label{fig:rearec} 
    \end{subfigure}
    \vspace{-0.1in}
    \caption{Visualization of Reasoning Trajectory} 
    \label{figure:case}
\end{figure}

\begin{figure}[h]
    \centering
    \includegraphics[width=0.45\textwidth]{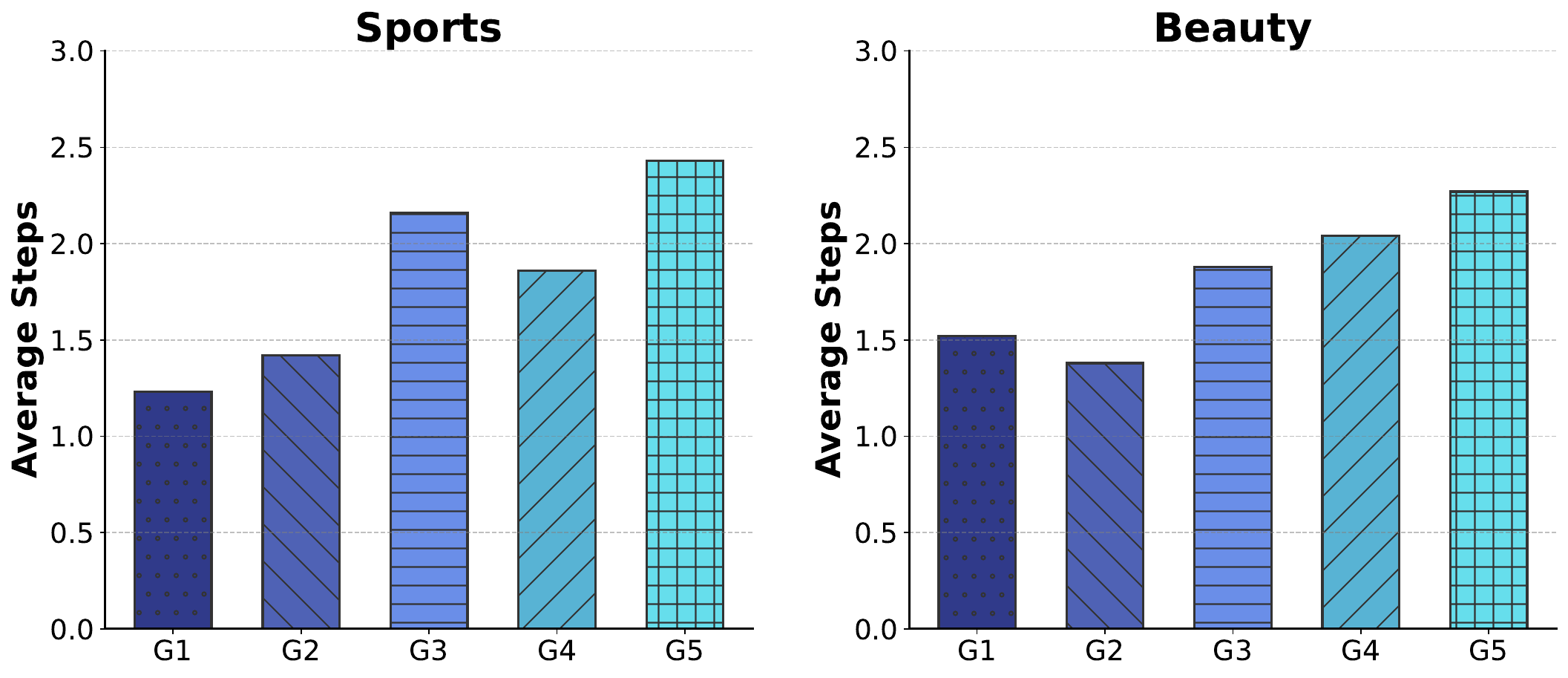}
    \vspace{-0.15in}
    \caption{Average reasoning steps for users of different levels of interactive length. G1 denotes the group of users with the lowest average number of interactions.}
    \label{figure:step}
\end{figure}

\subsection{Further Analysis}
\subsubsection{Case Study on Reasoning Trajectory}
To intuitively understand how HPS guides the reasoning process, we visualize the reasoning trajectories of DTRec and ReaRec using t-SNE in Figure~\ref{figure:case}. DTRec's trajectory progressively advances toward the target item, while ReaRec's trajectory is confined near the target from the outset. This is because ReaRec adopts overly strong process supervision, which forces each step to predict the final target directly, traping the reasoning process in a local optimum.

\subsubsection{Analysis of Adaptive Reasoning Depth}
We investigate how the ARH adapts to sequence complexity. Figure~\ref{figure:step} shows the average reasoning steps for sequences of varying lengths. The results indicate that the number of reasoning steps generally increases with sequence length. As longer sequences generally represent more complex user interests, the proposed ARH efficiently allocates more computational resources to complex patterns.

\section{Conclusion}
In this paper, we introduce DTRec, a framework that addresses the limitations of static reasoning in sequential recommendation. By incorporating Hierarchical Process Supervision for coarse-to-fine direction guidance and Adaptive Reasoning Halting for dynamic depth control, DTRec learns more effective and efficient reasoning trajectories.

\bibliographystyle{ACM-Reference-Format}
\bibliography{ref}

@inproceedings{kang2018self,
  title={Self-attentive sequential recommendation},
  author={Kang, Wang-Cheng and McAuley, Julian},
  booktitle={2018 IEEE international conference on data mining (ICDM)},
  pages={197--206},
  year={2018},
  organization={IEEE}
}

@inproceedings{sun2019bert4rec,
  title={BERT4Rec: Sequential recommendation with bidirectional encoder representations from transformer},
  author={Sun, Fei and Liu, Jun and Wu, Jian and Pei, Changhua and Lin, Xiao and Ou, Wenwu and Jiang, Peng},
  booktitle={Proceedings of the 28th ACM international conference on information and knowledge management},
  pages={1441--1450},
  year={2019}
}

@misc{tang2025thinkrecommendunleashinglatent,
  title={Think Before Recommend: Unleashing the Latent Reasoning Power for Sequential Recommendation}, 
  author={Jiakai Tang and Sunhao Dai and Teng Shi and Jun Xu and Xu Chen and Wen Chen and Wu Jian and Yuning Jiang},
  year={2025},
  eprint={2503.22675},
  archivePrefix={arXiv},
  primaryClass={cs.IR},
  url={https://arxiv.org/abs/2503.22675}, 
}

@misc{liu2025lareslatentreasoningsequential,
  title={LARES: Latent Reasoning for Sequential Recommendation}, 
  author={Enze Liu and Bowen Zheng and Xiaolei Wang and Wayne Xin Zhao and Jinpeng Wang and Sheng Chen and Ji-Rong Wen},
  year={2025},
  eprint={2505.16865},
  archivePrefix={arXiv},
  primaryClass={cs.IR},
  url={https://arxiv.org/abs/2505.16865}, 
}

@inproceedings{cen2020controllable,
  title={Controllable multi-interest framework for recommendation},
  author={Cen, Yukuo and Zhang, Jianwei and Zou, Xu and Zhou, Chang and Yang, Hongxia and Tang, Jie},
  booktitle={Proceedings of the 26th ACM SIGKDD international conference on knowledge discovery \& data mining},
  pages={2942--2951},
  year={2020}
}

@article{navon1977forest,
  title={Forest before trees: The precedence of global features in visual perception},
  author={Navon, David},
  journal={Cognitive psychology},
  volume={9},
  number={3},
  pages={353--383},
  year={1977},
  publisher={Elsevier}
}

@misc{zhou2024equivariantcontrastivelearningsequential,
  title={Equivariant Contrastive Learning for Sequential Recommendation}, 
  author={Peilin Zhou and Jingqi Gao and Yueqi Xie and Qichen Ye and Yining Hua and Jae Boum Kim and Shoujin Wang and Sunghun Kim},
  year={2024},
  eprint={2211.05290},
  archivePrefix={arXiv},
  primaryClass={cs.IR},
  url={https://arxiv.org/abs/2211.05290}, 
}

@article{hidasi2015session,
  title={Session-based recommendations with recurrent neural networks},
  author={Hidasi, Bal{\'a}zs and Karatzoglou, Alexandros and Baltrunas, Linas and Tikk, Domonkos},
  journal={arXiv preprint arXiv:1511.06939},
  year={2015}
}

\end{document}